\documentclass[12pt]{article}

\usepackage{color}
\usepackage{array}
\usepackage{epsfig}
\usepackage{amssymb}
\usepackage{graphics,graphpap}
\usepackage{subfigure}
\usepackage{amsmath}

\newcommand{\bal}{\begin{align}}

\def\beq{\begin{equation}}
\def\eeq{\end{equation}}
\def\bea{\begin{eqnarray}}
\def\eea{\end{eqnarray}}
\def\nn{\nonumber}
\def\nl{\nonumber\\}

\def\roughly#1{\mathrel{\raise.3ex\hbox
{$#1$\kern-.75em\lower1ex\hbox{$\sim$}}}}

\def\lesssim{\mathrel{\hbox{\rlap{\hbox{\lower4pt\hbox{$\sim$}}}\hbox{$<$}}}}
\def\gtrsim{\mathrel{\hbox{\rlap{\hbox{\lower4pt\hbox{$\sim$}}}\hbox{$>$}}}}

\def\sla#1{\raise.15ex\hbox{$/$}\kern-.57em #1}

\def\ket#1{\left| #1\right\rangle}
\def\ks{K_S}

\newcommand{\ba}{\begin{array}}
\newcommand{\ea}{\end{array}}

\def\bd{B_d^0}
\def\bs{B_s^0}
\def\bdbar{{\bar B}_d^0}
\def\bsbar{{\bar B}_s^0}
\def\btod{{\bar b} \to {\bar d}}
\def\btos{{\bar b} \to {\bar s}}
\def\Bsdecay{\bs\to J/\psi \phi}

\allowdisplaybreaks
\begin{document}

\begin{flushright}  
UMISS-HEP-2012-07\\
UdeM-GPP-TH-12-214 \\
\end{flushright}

\begin{center}
\bigskip
{\Large \bf \boldmath Reducing Penguin Pollution} \\

\bigskip
{\large 
Bhubanjyoti Bhattacharya $^{a,}$\footnote{bhujyo@lps.umontreal.ca},
Alakabha Datta $^{b,}$\footnote{datta@phy.olemiss.edu} \\
and David London $^{a,}$\footnote{london@lps.umontreal.ca}}
\\
\end{center}

\begin{flushleft}
~~~~~~~~~~~$a$: {\it Physique des Particules, Universit\'e
de Montr\'eal,}\\
~~~~~~~~~~~~~~~{\it C.P. 6128, succ.\ centre-ville, Montr\'eal, QC,
Canada H3C 3J7}\\
~~~~~~~~~~~$b$: {\it Department of Physics and Astronomy, 108 Lewis Hall, }\\ 
~~~~~~~~~~~~~~~{\it University of Mississippi, Oxford, MS 38677-1848, USA}\\
\end{flushleft}
\begin{center} 
\vskip0.5cm
{\Large Abstract\\}
\vskip3truemm

\parbox[t]{\textwidth} {The most common decay used for measuring
  $2\beta_s$, the phase of $\bs$-$\bsbar$ mixing, is $\Bsdecay$. This
  decay is dominated by the colour-suppressed tree diagram, but there
  are other contributions due to gluonic and electroweak penguin
  diagrams. These are often referred to as ``penguin pollution'' (PP)
  because their inclusion in the amplitude leads to a theoretical
  error in the extraction of $2\beta_s$ from the data. In the standard
  model (SM), it is estimated that the PP is negligible, but there is
  some uncertainty as to its exact size.  Now, $\phi_s^{c\bar{c}s}$
  (the measured value of $2\beta_s$) is small, in agreement with the
  SM, but still has significant experimental errors. When these are
  reduced, if one hopes to be able to see clear evidence of new
  physics (NP), it is crucial to have the theoretical error under
  control. In this paper, we show that, using a modification of the
  angular analysis currently used to measure $\phi_s^{c\bar{c}s}$ in
  $\Bsdecay$, one can reduce the theoretical error due to
  PP. Theoretical input is still required, but it is much more modest
  than entirely neglecting the PP.  If $\phi_s^{c\bar{c}s}$ differs
  from the SM prediction, this points to NP in the mixing.  There is
  also enough information to test for NP in the decay.  This method
  can be applied to all $\bs/\bsbar \to V_1 V_2$ decays.}

\end{center}

\bigskip
\leftline{Keywords: $B$ decays, CP violation, penguin pollution, angular analysis}
\bigskip
\leftline{PACS numbers: 11.30.Hv, 12.15.Ji, 13.25.Hw, 14.40.Nd}

\baselineskip=14pt

\newpage

\section{Introduction}

In the presence of $\bs$-$\bsbar$ mixing, the mass eigenstates
$B_{L,H}^s$ [$L$ ($H$) corresponds to ``light'' (``heavy'')] are
admixtures of the flavour eigenstates $\bs$ and $\bsbar$:
\beq
\ket{B_{L,H}^s} = p \ket{\bs} \pm q \ket{\bsbar} ~,
\eeq
where the complex coefficients $p$ and $q$ satisfy $|p|^2 + |q|^2 =1$.
States which are $\bs$ or $\bsbar$ at $t=0$ then evolve in time into
an admixture of both states, leading to the time-dependent states
$\bs(t)$ and $\bsbar(t)$. If both $\bs$ and $\bsbar$ can decay to the
final state $f$, there is an indirect (mixing-induced) CP-violating
asymmetry (CPA) between the rates $|\bs(t) \to f|^2$ and $|\bsbar(t)
\to f|^2$. The indirect CPA measures
\beq
{\rm Im}\left( \frac{q}{p} \frac{{\bar {\cal A}}_s^f}{{\cal A}_s^f} \right) ~,
\eeq
where ${\cal A}_s^f$ and ${\bar {\cal A}}_s^f$ are the amplitudes for
$\bs \to f$ and $\bsbar \to f$, respectively.  ${\bar {\cal A}}_s^f$
can be obtained from ${\cal A}_s^f$ by changing the signs of the weak
phases.  

$\bs$-$\bsbar$ mixing is dominated by the box diagram with an internal
$t$ quark, so that $q/p = (V_{tb}^* V_{ts}/V_{tb} V_{ts}^*) = {\rm
  exp}(2i \arg(V_{tb}^* V_{ts}))$. This is phase-convention
dependent. Suppose now that ${\cal A}_s^f$ is dominated by a single
decay amplitude with the Cabibbo-Kobayashi-Maskawa (CKM) matrix
elements $V_{cb}^* V_{cs}$. We then have ${\bar {\cal A}}_s^f/{\cal
  A}_s^f = (V_{cb} V_{cs}^*/V_{cb}^* V_{cs}) = {\rm exp}(2i
\arg(V_{cb} V_{cs}^*))$, which is also phase-convention
dependent. However, the product of these two quantities is
\beq
\frac{q}{p} \frac{{\bar {\cal A}}_s^f}{{\cal A}_s^f} = \frac{V_{tb}^*
  V_{ts}}{V_{cb}^* V_{cs}} \frac{V_{cb} V_{cs}^*}{V_{tb} V_{ts}^*} =
e^{2i \beta_s} ~,
\eeq
where
\beq
\label{betasdef}
\beta_s \equiv \arg \left[ -\frac{V_{tb}^* V_{ts}}{V_{cb}^* V_{cs}} \right] ~.
\eeq
This is phase-convention independent, and hence physical. The indirect
CPA measures $\sin 2\beta_s$.

In the standard model (SM), $2\beta_s$ is expected to be very small:
the SM prediction is $2\beta_s = 0.03636 \pm 0.00170~{\rm rad}$
\cite{Lenzetal}. For this reason there is great interest in measuring
this quantity. A value for $2\beta_s$ significantly different from
zero would be a smoking-gun signal of new physics (NP).

The main decay mode used for measuring $2\beta_s$ is $\Bsdecay$, which
is the analogue of the ``golden'' mode $\bd\to J/\psi \ks$ used to
measure the $\bd$-$\bdbar$ mixing phase $2\beta$. Since $\Bsdecay$
involves two final-state vector mesons, a time-dependent angular
analysis of this decay (with $J/\psi \to \ell^+ \ell^-$ and $\phi\to
K^+ K^-$) must be used to disentangle the CP $+$ and $-$ final states.
The most precise measurement of the indirect CPA has been performed by
the LHCb Collaboration \cite{LHCbBsmixing}. They find\footnote{
  $\beta_s$ is defined in Eq.~(\ref{betasdef}). For the measured
  $\bs$-$\bsbar$ mixing phase, it is common to use the symbol
  $\phi_s^{c\bar{c}s}$ (or occasionally just $\phi_s$), which is equal
  to $-2\beta_s$ in the SM.}
\beq
\phi_s^{c\bar{c}s} = -0.001 \pm 0.101~({\rm stat}) \pm 0.027~({\rm
  syst})~{\rm rad} ~,
\label{2betasmeas}
\eeq
in agreement with the SM. (Still, the errors are large enough that NP
cannot be excluded.)

Now, the above formalism holds for the case where ${\cal A}_s^f$ is
dominated by a single decay amplitude.  However, if ${\cal A}_s^f$
contains a second decay amplitude with a different weak phase, then
the indirect CPA no longer cleanly measures $2\beta_s$ -- there is a
theoretical uncertainty due to the presence of the second
amplitude. It is often the case that the first and second amplitudes
correspond to tree and penguin diagrams, respectively. The theoretical
uncertainty is thus generally called ``penguin pollution.''

For $\Bsdecay$, the decay amplitude can be written
\beq
{\cal A}_s^{J/\psi \phi} = \lambda^{(s)}_c C' + (\lambda^{(s)}_t P'_t
+ \lambda^{(s)}_c P'_c + \lambda^{(s)}_u P'_u)
+ \frac23 \lambda^{(s)}_t P'_{EW} ~,
\eeq
where $\lambda^{(q')}_q \equiv V_{qb}^* V_{qq'}$.  Here, $C'$, $P'$
and $P'_{EW}$ are the colour-suppressed tree, gluonic penguin and
electroweak penguin diagrams, respectively.  (As this is a $\btos$
transition, the diagrams are written with primes.) The index `$i$' of
$P'_i$ indicates the flavour of the quark in the loop. The unitarity
of the CKM matrix ($\lambda^{(s)}_u + \lambda^{(s)}_c +
\lambda^{(s)}_t = 0$) can be used to eliminate the $t$-quark
contribution:
\bea
\label{ABdefs}
{\cal A}_s^{J/\psi \phi} & = & \lambda^{(s)}_c (C' + P'_{ct} - \frac23 P'_{EW}) + \lambda^{(s)}_u (P'_{ut} - \frac23 P'_{EW}) \nn\\
&\equiv& e^{i \arg(V_{cb}^* V_{cs})} \left[A_1 + e^{i\gamma} A_2 \right]~,
\eea
where $P'_{ct} \equiv P'_c - P'_t$, $P'_{ut} \equiv P'_u - P'_t$. In
the second line, $A_1$ and $A_2$ are combinations of diagrams, and
include the magnitudes of the corresponding CKM matrix elements.  We
have explicitly written the weak-phase dependence of the terms in the
amplitude\footnote{The weak phase in the coefficient of $A_2$ is
  $\arg(V_{ub}^* V_{us} / V_{cb}^* V_{cs})$, and so should technically
  be called $\gamma_s$. However, it differs very little from $\gamma
  \equiv \arg(V_{ub}^* V_{ud} / V_{cb}^* V_{cd})$, and so we refer to
  it by this name.}.  (The phase information in the CKM matrix is
conventionally parametrized in terms of the unitarity triangle, in
which the interior (CP-violating) angles are known as $\alpha$,
$\beta$ and $\gamma$ \cite{pdg}.) $A_1$ and $A_2$ contain strong
phases. 

We now have
\beq
\label{phasecombination}
\frac{q}{p} \frac{{\bar {\cal A}}_s^f}{{\cal A}_s^f} = e^{2i\beta_s} \frac{A_1 + e^{-i\gamma} A_2}{A_1 + e^{i\gamma} A_2} ~.
\eeq
The point here is that the $A_2$ term represents the penguin pollution
(PP).  There are several ingredients in estimating its size. First,
$|V_{ub}^* V_{us}|/|V_{cb}^* V_{cs}| \approx 0.02$. Second, in
Ref.~\cite{GHLR1,GHLR2}, it is estimated that $|P'_{ct}/C'| \sim |P'_{ut}/C'|
= O(1)$. However, here the penguin diagrams are both OZI-suppressed,
so these ratios are much smaller. Third, it is expected that
$|P'_{EW}/C'| = O({\bar\lambda})$, where ${\bar\lambda} \sim 0.2$
\cite{GHLR1,GHLR2}. Putting everything together, it is estimated
theoretically that $|A_2 / A_1| = O(10^{-3})$ \cite{smallcorr1,smallcorr2,smallcorr3}. Thus,
to a good approximation, the $A_2$ term, i.e.\ the PP, is negligible.
In this case, the indirect CPA in $\Bsdecay$ cleanly measures
$2\beta_s$.

Indeed, in the determination of $\phi_s^{c\bar{c}s}$
[Eq.~(\ref{2betasmeas})], the LHCb Collaboration neglected the PP
\cite{LHCbBsmixing} (${\bar {\cal A}}_s^f = {\cal A}_s^f$ was
assumed).  However, the result reintroduces the question of the size
of the PP.  To be specific, we now know that $\phi_s^{c\bar{c}s}$ is
small.  Suppose a future measurement finds $\phi_s^{c\bar{c}s} = 0.1
\pm 0.01~{\rm rad}$ (error is statistical only).  This disagrees with
the SM prediction and therefore suggests NP. But perhaps the size of
the PP has been underestimated, so that there is really a nonzero
theoretical error associated with this measurement.  It is possible
that, when this error is taken into account, the discrepancy with the
SM disappears, so that there is no signal of NP.  This situation is
not at all improbable -- the theoretical prediction of $O(10^{-3})$
for the size of the PP includes estimates of the hadronic matrix
elements. But these are notoriously difficult to determine with
certainty. It is not impossible that the true size of the PP is larger
than its theoretical estimate, and that this introduces a theoretical
error which is important when the measured value of
$\phi_s^{c\bar{c}s}$ is small.

This issue was raised several years ago in Ref.~\cite{JpsiK*}. There
it was suggested that the PP term can be measured in the
time-dependent angular distribution of the flavour-specific $\btod$
decay $\bs\to J/\psi {\bar K}^{*0} (\to \pi^+ K^-)$, and then related
to $\Bsdecay$ using flavour SU(3) symmetry. The difficulty here is
that the value of SU(3) breaking is unknown, and this is problematic
given that we need a precise value of the size of the PP.

The purpose of this paper is to point out that, in fact, using the
present time-dependent angular analysis, the theoretical error due to
PP associated with $\phi_s^{c\bar{c}s}$ can be reduced. In this
method, one retains the PP term. One piece of theoretical input is
still needed to extract $\phi_s^{c\bar{c}s}$, but the theoretical
error is quite a bit smaller than when the PP term is neglected from
the beginning. In addition, if it is concluded that NP is present,
there is enough measured information to determine whether the NP is in
the mixing and/or the decay.  

In Sec.~2, we describe the method for extracting $\phi_s^{c\bar{c}s}$
from the full angular analysis.  Here we follow the procedure as
described in Ref.~\cite{Psiphibetas}. The theoretical error, and how
it can be reduced, are discussed in Sec.~3. In Sec.~4, we show how the
PP or NP parameters can be extracted, and discuss the application of
the method to other $\bs/\bsbar \to V_1 V_2$ decays. We conclude in
Sec.~5.

\section{Angular Analysis}

Given that $\Bsdecay$ has two final-state vector mesons, its amplitude
can be separated into 3 helicities: the polarizations of the vector
mesons are either longitudinal ($A_0$), or transverse to their
directions of motion and parallel ($A_\|$) or perpendicular
($A_\perp$) to one another. In addition, since the $\phi$ is detected
through its decay to $K^+ K^-$, one must also allow for the
possibility that the observed $K^+ K^-$ pair has relative angular
momentum $l=0$ ($S$-wave) \cite{Swave}. That is, the angular
distribution of the decay $\bs \to J/\psi (\to \ell^+ \ell^-) \phi
(\to K^+ K^-)$ involves the 4 helicities $0$, $\|$, $\perp$, $S$. In
going from the decay to the CP-conjugate decay, $A_h \to \eta_h {\bar
  A}_h$ ($h= 0, \|, \perp, S$), in which the ${\bar A}_h$ are equal
to the $A_h$, but with weak phases of opposite sign, and $\eta_0 =
\eta_\| = +1$, $\eta_\perp = \eta_S = - 1$.

The angular distribution is given in terms of the vector $\vec{\Omega}
= (\theta, \psi, \varphi)$ in the transversity basis
\cite{DDLR1,DDLR2,DDLR3,DFN,CW1,CW2}. This basis is defined as follows: in the $J/\psi$
rest frame, one has a righthanded coordinate system in which the $x$
axis is parallel to ${\vec p}_\phi$ and the $z$ axis is parallel to
${\vec p}_{K^-} \times {\vec p}_{K^+}$. In this frame, $\theta$ and
$\varphi$ are the azimuthal and polar angles, respectively, of the
$\mu^+$. The angle $\psi$ is the angle between ${\vec p}_{K^-}$ and
${\vec p}_{J/\psi}$ in the rest frame of the $\phi$. We have
\bea
\label{AD:Bsphiphi}
\frac{d^4 \Gamma(\Bsdecay)}{dt d\vec{\Omega}} &\propto& \sum^{10}_{k=1} h_k(t) f_k(\vec{\Omega}) ~.
\eea
The explicit expressions for the $f_k(\vec{\Omega})$ are given in
Ref.~\cite{Psiphibetas}.

The $h_k(t)$ are determined as follows.  The time dependence of the
helicity amplitudes $A_h$ ($h=0, \|, \perp, S$) is given by
\cite{Datta:2012ky}
\bea
\label{eq:3}
A_h(t) &=& g_+(t) A_h(t = 0) + \eta_h \, \frac{q}{p} \, g_-(t) {\bar A}_h(t = 0)~.
\eea
$g_{\pm}(t)$ contain the information about the time evolution of
$\bs(t)$ and $\bsbar(t)$. We use the following established relations
involving $g_\pm(t)$:
\bea
\label{gplusminus}
|g_\pm(t)|^2 &=& \frac12 e^{ - \Gamma_s t} \Big(\cosh{(\Delta \Gamma_s/2) t}\,\pm \cos{ \Delta m_s t}\Big)  ~, \nl
g^*_+(t) g_-(t) &=& \frac12 e^{ - \Gamma t} \Big(-\sinh{(\Delta \Gamma_s/2) t}\,+ i  \sin{ \Delta m_s t}\Big) ~.
\eea
The $h_k(t)$ are given by
\bea
\label{eq:2}
h_1(t)=|A_0(t)|^2 ~,~~ h_2(t)=|A_\|(t)|^2 &,~& h_3(t)=|A_\perp(t)|^2
~,~~h_7(t)=|A_S(t)|^2 ~, \nn\\
h_4(t)={\rm Im}\left( A_\perp(t)A_\|^*(t)\right) &,~& h_6(t)={\rm
  Im}\left( A_\perp(t)A_0^*(t)\right) ~, \nn\\
h_5(t)={\rm Re}\left( A_0(t)A_\|^*(t)\right) &,~& h_9(t) ={\rm Im}\left(
A_\perp(t) A_S^*(t)\right) ~, \nn \\
h_8(t)={\rm
  Re}\left( A_\|(t) A_S^*(t)\right) &,~& h_{10} (t)={\rm Re}\left(
A_0(t) A_S^*(t)\right) ~,
\eea
and can be written as
\bea
\label{abcddefs}
h_k(t) & = & \frac12 e^{ - \Gamma_s t} \left[ c_k \cos{ \Delta m_s t} + d_k \sin{ \Delta m_s t} \right. \nl
&& \hskip2truecm \left. +~a_k \cosh{(\Delta \Gamma_s/2) t} + b_k \sinh{(\Delta \Gamma_s/2) t} \right] ~.
\eea
By measuring the time-dependent angular distribution and fitting to
the four time-dependent functions, $\Gamma_s$ and $\Delta \Gamma_s$
can be determined, as well as the coefficients $a_k$-$d_k$. (Note that
the width difference in the $B^0_q$-${\bar B}^0_q$ system, $\Delta
\Gamma_q$, is sizeable only for $\bs$ decays. Thus, this method for
dealing with PP cannot be used for $\bd$ decays.)

There are also the functions ${\bar h}_k(t)$ in the angular
distribution of the CP-conjugate decay $\bsbar\to J/\psi \phi$.  They
can be obtained from the $h_k(t)$ by making the following changes:
$A_h \leftrightarrow \eta_h {\bar A}_h$ and $\phi_s^{c\bar{c}s} \to
-\phi_s^{c\bar{c}s}$. Each of the ${\bar a}_k$-${\bar d}_k$ of the
${\bar h}_k(t)$ is the same as the $a_k$-$d_k$ of the $h_k(t)$, up to
a possible sign. Both $h_k(t)$ and ${\bar h}_k(t)$ should be included
in the fit.

Also, the $h_k(t)$ contain both CP-conserving and CP-violating
terms. However, when one calculates the sum or difference of $h_k(t)$
and ${\bar h}_k(t)$, certain terms cancel, so that these sums and
differences are purely CP-conserving or CP-violating. In particular,
the untagged sums $h_k(t) + {\bar h}_k(t)$ are CP-violating for
$k=4,6,8,10$. These correspond to triple-product terms. Similarly, the
differences $h_k(t) - {\bar h}_k(t)$ ($k=1,2,3,5,7,9$) are
CP-violating. These can only be constructed through tagged decays, and
contain the direct and indirect CPA's.

Now, the ultimate goal is to measure $\phi_s^{c\bar{c}s}$. To this
end, it is necessary to express the $a_k$-$d_k$ in terms of all the
theoretical unknowns in order to determine what exactly can be
extracted. The $b_k$ and $d_k$ involve terms of the form
\beq
\label{phaseconvention}
{\rm Re} \left( \frac{q}{p} A_h^* {\bar A}_{h'} \right) ~~,~~~~
{\rm Im} \left( \frac{q}{p} A_h^* {\bar A}_{h'} \right) ~.
\eeq
The helicity amplitudes can be multiplied by an arbitrary phase, so
that the form of the individual terms is uncertain.  In order to fix
the phases, in what follows we adopt the convention that the dominant
SM decay amplitude has no weak phase. That is, the phase term $e^{-2i
  \arg(V_{cb}^* V_{cs})}$ is factored out and combined with the phase
of $q/p$ to make $2\beta_s$, as in Eq.~(\ref{phasecombination}).
Thus, any terms of the form ${\rm Re} ( q/p )$ or ${\rm Im} ( q/p )$
depend only on the mixing phase $\beta_s$.

The $a_k$-$d_k$ are expressed in terms of the unknown parameters.
Suppose first that only the dominant SM amplitude is retained (the
``1-amplitude method'') -- this is $A_1$ of Eq.~(\ref{ABdefs}). In
this case we have ${\bar A}_h = A_h$ ($h=0,\|,\perp,S$), so that there
are 8 unknown parameters -- the magnitudes of the $A_h$ (4), the
relative strong phases (3), and $\phi_s^{c\bar{c}s}$. (In their
analysis, LHCb fix the normalization $|A_0|^2 + |A_\||^2 + |A_\perp|^2
= 1$.)  The coefficients $a_k$-$d_k$ are given in Table
\ref{abcdtableA1}, in which the strong-phase differences are
$\delta_{ij} \equiv \arg (A_i) - \arg (A_j)$. It is clear from this
Table that all unknown parameters can be determined from the
measurements of $a_k$-$d_k$. In addition, there is a great deal of
redundancy, which allows for a reasonably precise determination of
these unknowns. Indeed, this is the method used by the LHCb
Collaboration \cite{LHCbBsmixing,Psiphibetas}. However, as noted
above, there is a potential theoretical error associated with the
neglect of the PP.

\begin{table}[tbh]
\center
{\hskip-2truecm
\small
\begin{tabular}{ccccc}
\hline
\hline
& $a_k$ & $b_k$ & $c_k$ & $d_k$ \\ \hline
$h_1$ & $2|A_0|^2$ & $-2|A_0|^2 \cos \phi_s^{c\bar{c}s}$ & 0 & $2|A_0|^2 \sin \phi_s^{c\bar{c}s}$ \\
$h_2$ & $2|A_\||^2$ & $-2|A_\||^2 \cos \phi_s^{c\bar{c}s}$ & 0 & $2|A_\||^2 \sin \phi_s^{c\bar{c}s}$ \\
$h_3$ & $2|A_\perp|^2$ & $2|A_\perp|^2 \cos \phi_s^{c\bar{c}s}$ & 0 & $-2 |A_\perp|^2 \sin \phi_s^{c\bar{c}s}$ \\
$h_4$ & 0 & $-2 |A_\perp||A_\||\cos\delta_{\perp\|} \sin \phi_s^{c\bar{c}s}$ & $2|A_\perp||A_\||\sin\delta_{\perp\|}$ & $-2 |A_\perp||A_\||\cos\delta_{\perp\|} \cos \phi_s^{c\bar{c}s}$ \\
$h_5$ & $2|A_\|||A_0|\cos\delta_{\|0}$ & $-2|A_\|||A_0|\cos\delta_{\|0} \cos \phi_s^{c\bar{c}s}$ & 0 & $2 |A_\|||A_0|\cos\delta_{\|0} \sin \phi_s^{c\bar{c}s}$ \\
$h_6$ & 0 & $-2 |A_\perp||A_0|\cos\delta_{\perp 0} \sin \phi_s^{c\bar{c}s}$ & $2|A_\perp||A_0|\sin\delta_{\perp0}$ & $-2 |A_\perp||A_0|\cos\delta_{\perp 0} \cos \phi_s^{c\bar{c}s}$ \\
$h_7$ & $2|A_S|^2$ & $2|A_S|^2 \cos \phi_s^{c\bar{c}s} $ & 0 & $-2|A_S|^2 \sin \phi_s^{c\bar{c}s}$ \\
$h_8$ & 0 & $-2 |A_\|||A_S|\sin\delta_{\|S} \sin \phi_s^{c\bar{c}s}$ & $2|A_\|||A_S|\cos\delta_{\|S}$ & $-2 |A_\|||A_S|\sin\delta_{\|S} \cos \phi_s^{c\bar{c}s} $ \\
$h_9$ & $2|A_\perp||A_S|\sin\delta_{\perp S}$ & $2 |A_\perp||A_S|\sin\delta_{\perp S} \cos \phi_s^{c\bar{c}s}$ & 0 & $- 2 |A_\perp||A_S|\sin\delta_{\perp S} \sin \phi_s^{c\bar{c}s}$ \\
$h_{10}$ & 0 & $-2 |A_0||A_S|\sin\delta_{0S} \sin \phi_s^{c\bar{c}s}$ & $2|A_0||A_S|\cos\delta_{0S}$ & $-2 |A_0||A_S|\sin\delta_{0S} \cos \phi_s^{c\bar{c}s} $ \\
\hline
\hline
\end{tabular}
}
\caption{Coefficients $a_k$-$d_k$ of Eq.~(\ref{abcddefs}) in terms of
  $|A_h|$ ($h=0,\|,\perp,S$) and $\delta_{ij}$ [Eq.~(\ref{phasedefs})]
  for the case where ${\bar A}_h = A_h$.}
\label{abcdtableA1}
\end{table}

Suppose instead that no assumptions about the decay amplitude are
made, and we allow for the possibility of more than one contribution
to the amplitude. In fact, this was first considered in
Ref.~\cite{Fleischer99}. Here (and in Ref.~\cite{JpsiK*}), the focus
was specifically on the SM. The PP term was included, and the tree
($A_1$) and penguin ($A_2$) amplitudes were separated, as in
Eq.~(\ref{ABdefs}). It was then noted that, if one takes the mixing
phase as known, the angular analysis gives enough information to
extract $\gamma$, given one piece of theoretical input. This is
similar to one of the points made later in Sec.~4. However, the
possibility of NP was not considered. Indeed, in the presence of NP,
the method of Ref.~\cite{Fleischer99} no longer holds. On the other
hand, as we show below, one can go well beyond this analysis by {\it
  not} separating $A_1$ and $A_2$. By applying the angular analysis to
the full amplitudes $A_h$ and ${\bar A}_h$, one can still extract
$\phi_s^{c\bar{c}s}$, even if there is NP in the PP. Furthermore, we
show how the theoretical error due to the PP can be reduced in this
way.

The most general way to allow for the possibility of more than one
contribution to the amplitude (the ``2-amplitude method'') is to
consider the full amplitudes $A_h$ and ${\bar A}_h$, and take ${\bar
  A}_h \ne A_h$. This inequality can arise due to enhanced PP within
the SM and/or the presence of NP. In general, the $a_k$-$d_k$ are
expressed in terms of 16 unknown parameters: the magnitudes of the
$A_h$ and ${\bar A}_h$ (8), their relative phases (7), and
$\phi_s^{c\bar{c}s}$. For the phase differences we define
\bea
\label{phasedefs}
\delta_{ij} &\equiv& \arg (A_i) - \arg (A_j) ~, \nn\\
\bar{\delta}_{ij} &\equiv& \arg (\bar{A}_i) - \arg (\bar{A}_j) ~, \nn\\
D_{ij} &\equiv& \arg (\bar{A}_i) - \arg (A_j) ~,
\eea
where $i,j$ are any of the 4 helicities $0,\|,\perp,S$. Of course,
there are many phase differences, but only 7 are independent. It is
straightforward to show that
\bea
\label{Drels}
D_{ij} + D_{ji} &=& D_{ii} + D_{jj} ~, \nn\\
D_{ii} - D_{jj} &=& \bar{\delta}_{ij} - \delta_{ij} ~.
\eea

When ${\bar A}_h \ne A_h$, the coefficients $a_k$-$d_k$ of Table
\ref{abcdtableA1} are modified -- in particular, the $a_k$ and $c_k$
which vanish in Table \ref{abcdtableA1} are now nonzero.  The
$a_k$-$d_k$ for this case are given in Tables
\ref{akcktable}-\ref{dktable}. All of these quantities can be measured
in the time-dependent angular analysis.

\begin{table}[tbh]
\center
\begin{tabular}{ccc}
\hline
\hline
& $(a_k + c_k)/2$ & $(a_k - c_k)/2$ \\ \hline
$h_1$ & $|A_0|^2$ & $|{\bar A}_0|^2$ \\
$h_2$ & $|A_\||^2$ & $|{\bar A}_\||^2$ \\
$h_3$ & $|A_\perp|^2$ & $|{\bar A}_\perp|^2$ \\
$h_4$ & $|A_\perp||A_\||\sin\delta_{\perp\|}$ & $-|\bar{A}_\perp||\bar{A}_\||\sin\bar{\delta}_{\perp\|}$ \\
$h_5$ & $|A_\|||A_0|\cos\delta_{\|0}$ & $|\bar{A}_\|||\bar{A}_0|\cos\bar{\delta}_{\|0}$ \\
$h_6$ & $|A_\perp||A_0|\sin\delta_{\perp0}$ & $-|\bar{A}_\perp||\bar{A}_0|\sin\bar{\delta}_{\perp0}$ \\
$h_7$ & $|A_S|^2$ & $|{\bar A}_S|^2$ \\
$h_8$ & $|A_\|||A_S|\cos\delta_{\|S}$ & $-|\bar{A}_\|||\bar{A}_S|\cos\bar{\delta}_{\|S}$ \\
$h_9$ & $|A_\perp||A_S|\sin\delta_{\perp S}$ & $|\bar{A}_\perp||\bar{A}_S|\sin\bar{\delta}_{\perp S}$ \\
$h_{10}$ & $|A_0||A_S|\cos\delta_{0S}$ & $-|\bar{A}_0||\bar{A}_S|\cos\bar{\delta}_{0S}$ \\
\hline
\hline
\end{tabular}
\caption{Coefficients $a_k$ and $c_k$ of Eq.~(\ref{abcddefs}) in terms
  of $|A_h|$ and $|{\bar A}_h|$ ($h=0,\|,\perp,S$), and $\delta_{ij}$
  and $\bar{\delta}_{ij}$ [Eq.~(\ref{phasedefs})].}
\label{akcktable}
\end{table}

\begin{table}[tbh]
\center
\begin{tabular}{cc}
\hline
\hline
& $b_k$ \\ \hline
$h_1$ & $-2|A_0||\bar{A}_0|\cos(\phi_s^{c\bar{c}s} - D_{00})$ \\
$h_2$ & $-2|A_\|||\bar{A}_\||\cos(\phi_s^{c\bar{c}s} - D_{\|\|})$ \\
$h_3$ & $2|A_\perp||\bar{A}_\perp|\cos(\phi_s^{c\bar{c}s} - D_{\perp\perp})$ \\
$h_4$ & $-|A_\perp||\bar{A}_\||\sin(\phi_s^{c\bar{c}s} - D_{\|\perp})
- |\bar{A}_\perp||A_\||\sin(\phi_s^{c\bar{c}s} - D_{\perp\|})$ \\
$h_5$ & $-|A_\|||\bar{A}_0|\cos(\phi_s^{c\bar{c}s} - D_{0\|})
- |\bar{A}_\|||A_0|\cos(\phi_s^{c\bar{c}s} - D_{\|0})$ \\
$h_6$ & $-|A_\perp||\bar{A}_0|\sin(\phi_s^{c\bar{c}s} - D_{0\perp})
- |\bar{A}_\perp||A_0|\sin(\phi_s^{c\bar{c}s} - D_{\perp 0})$ \\
$h_7$ & $2|A_S||\bar{A}_S|\cos(\phi_s^{c\bar{c}s} - D_{SS})$ \\
$h_8$ & $-|\bar{A}_\|||A_S|\cos(\phi_s^{c\bar{c}s} - D_{\|S})
+ |A_\|||\bar{A}_S|\cos(\phi_s^{c\bar{c}s} - D_{S\|})$ \\
$h_9$ & $-|\bar{A}_\perp||A_S|\sin(\phi_s^{c\bar{c}s} - D_{\perp S})
+ |A_\perp||\bar{A}_S|\sin(\phi_s^{c\bar{c}s} - D_{S\perp})$ \\
$h_{10}$ & $-|\bar{A}_0||A_S|\cos(\phi_s^{c\bar{c}s} - D_{0S})
+ |A_0||\bar{A}_S|\cos(\phi_s^{c\bar{c}s} - D_{S0})$ \\
\hline
\hline
\end{tabular}
\caption{Coefficients $b_k$ of Eq.~(\ref{abcddefs}) in terms of
  $|A_h|$ and $|{\bar A}_h|$ ($h=0,\|,\perp,S$), $D_{ij}$
  [Eq.~(\ref{phasedefs})], and $\phi_s^{c\bar{c}s}$.}
\label{bktable}
\end{table}

\begin{table}[tbh]
\center
\begin{tabular}{cc}
\hline
\hline
& $d_k$ \\ \hline
$h_1$ & $2|A_0||\bar{A}_0|\sin(\phi_s^{c\bar{c}s} - D_{00})$ \\
$h_2$ & $2|A_\|||\bar{A}_\||\sin(\phi_s^{c\bar{c}s} - D_{\|\|})$ \\
$h_3$ & $-2|A_\perp||\bar{A}_\perp|\sin(\phi_s^{c\bar{c}s} - D_{\perp\perp})$ \\
$h_4$ & $-|A_\perp||\bar{A}_\||\cos(\phi_s^{c\bar{c}s} - D_{\|\perp})
- |\bar{A}_\perp||A_\||\cos(\phi_s^{c\bar{c}s} - D_{\perp\|})$ \\
$h_5$ & $|A_\|||\bar{A}_0|\sin(\phi_s^{c\bar{c}s} - D_{0\|})
+ |\bar{A}_\|||A_0|\sin(\phi_s^{c\bar{c}s} - D_{\|0})$ \\
$h_6$ & $-|A_\perp||\bar{A}_0|\cos(\phi_s^{c\bar{c}s} - D_{0\perp})
- |\bar{A}_\perp||A_0|\cos(\phi_s^{c\bar{c}s} - D_{\perp0})$ \\
$h_7$ & $-2|A_S||\bar{A}_S|\sin(\phi_s^{c\bar{c}s} - D_{SS})$ \\
$h_8$ & $|\bar{A}_\|||A_S|\sin(\phi_s^{c\bar{c}s} - D_{\|S})
- |A_\|||\bar{A}_S|\sin(\phi_s^{c\bar{c}s} - D_{S\|})$ \\
$h_9$ & $-|\bar{A}_\perp||A_S|\cos(\phi_s^{c\bar{c}s} - D_{\perp S})
+ |A_\perp||\bar{A}_S|\cos(\phi_s^{c\bar{c}s} - D_{S\perp})$ \\
$h_{10}$ &  $|\bar{A}_0||A_S|\sin(\phi_s^{c\bar{c}s} - D_{0S})
- |A_0||\bar{A}_S|\sin(\phi_s^{c\bar{c}s} - D_{S0})$ \\
\hline
\hline
\end{tabular}
\caption{Coefficients $d_k$ of Eq.~(\ref{abcddefs}) in terms of
  $|A_h|$ and $|{\bar A}_h|$ ($h=0,\|,\perp,S$), $D_{ij}$
  [Eq.~(\ref{phasedefs})], and $\phi_s^{c\bar{c}s}$.}
\label{dktable}
\end{table}

$|A_h|$ and $|{\bar A}_h|$ can be obtained from $a_k$ and $c_k$ for
$k=1,2,3,7$ (Table \ref{akcktable}).  The 3 independent relative
phases among the $A_h$ (the $\delta_{ij}$) and the 3 independent
relative phases among the ${\bar A}_h$ (the $\bar{\delta}_{ij}$) can
be found from $a_k$ and $c_k$ for $k=4$-6 and 8-10.  For the missing
relative phase -- a $D_{ij}$ -- one has to look at the $b_k$ and $d_k$
(Tables \ref{bktable},\ref{dktable}). However, all terms in these
tables are a function of $\phi_s^{c\bar{c}s} - D_{ij}$.  Thus, it is
not possible to separate $\phi_s^{c\bar{c}s}$ and the $D_{ij}$ solely
from the data. Even in this case, where the PP has been retained,
theoretical input is necessary. (Note that this not an accident -- it
is a direct consequence of the fact that one requires a convention to
fix the phases in Eq.~(\ref{phaseconvention}).)

On the other hand, it is obvious how to formulate the theoretical
input -- one has to choose a value for one of the $D_{ij}$. For
example, suppose we set $D_{00} = 0$.  With this input, we can now
obtain $\phi_s^{c\bar{c}s}$ from a fit to the data. Note that no
assumptions are made about the magnitudes and other relative phases of
the $A_h$ and ${\bar A}_h$ -- they are also obtained from the fit.

This choice of input does raise two questions: why choose $D_{00}$ and
why set it equal to 0? The answer to both questions is as follows. The
best-fit values of $|A_h|$, $|{\bar A}_h|$, $\delta_{ij}$ and
$\bar{\delta}_{ij}$ are all independent of the choice of input. In
changing the input, $\phi_s^{c\bar{c}s}$ and the $D_{ij}$ all shift by
known quantities. For example, had we chosen, say, $D_{\|\|} = 0$, the
best-fit value of $\phi_s^{c\bar{c}s}$ would be shifted by $D_{\|\|} -
D_{00} = \bar\delta_{\|0} - \delta_{\|0}$ [Eq.~(\ref{Drels})], and the
values of the other $D_{ij}$ would also be changed according to
Eq.~(\ref{Drels}). And had we chosen, for example, $D_{00} = \pi/6$,
this would simply have the effect of shifting the best-fit values of
the $D_{ij}$ and $\phi_s^{c\bar{c}s}$ by $\pi/6$. Since, as we will
see in the next section, the $D_{ij}$ are all expected to be small, we
choose as input $D_{00} = 0$ (which is the same as in the 1-amplitude
method).

\section{Theoretical Error}

In the 2-amplitude method, the theoretical error on
$\phi_s^{c\bar{c}s}$ is equal to that associated with the assumption
that $D_{00} = 0$. In the SM we write
\beq
A_0 = A_{1,0} + e^{i\gamma} A_{2,0} ~,
\eeq
where the $A_{2,0}$ term represents the PP. Then
\beq
\label{D00eqn}
D_{00} = \arg(A^*_0\bar{A}_0) \simeq 2 \, \frac{|A_{2,0}|}{|A_{1,0}|} \, \cos\Delta \sin\gamma ~,
\eeq
where $\Delta$ is the strong-phase difference between $A_{2,0}$ and
$A_{1,0}$. This shows two things. First, the theoretical error
associated with the assumption that $D_{00} = 0$ is $O(|A_{2,0}|/|A_{1,0}|)$. Second, if
one wishes to calculate its exact value, one has to compute the ratio
$|A_{2,0}/A_{1,0}|$, which involves hadronic matrix elements, and one
also needs to determine the value of the strong phase $\Delta$.  The
upshot is that it is virtually impossible to reliably calculate the
size of the theoretical error associated with $D_{00} = 0$.

We therefore see that the 2-amplitude method requires theory input,
with an associated theoretical error of $O(|A_{2,0}|/|A_{1,0}|)$. At
first glance, this does not seem to be much of an improvement over the
1-amplitude method. However, let us examine the theory errors of both
methods in more detail. While the 2-amplitude method requires one
assumption -- $D_{00} = 0$ -- the 1-amplitude method in fact requires
8 assumptions:
\beq
|{\bar A}_h| = |A_h| ~(h=0,\|,\perp,S) ~,~~ \bar{\delta}_{ij} = \delta_{ij} ~(ij = \|0,\perp 0,0S) ~,~~ D_{00} = 0 ~.
\label{assumptions}
\eeq
In general, there is a theoretical error associated with each
assumption. This is due to the fact that, if the true values of the
parameters do not obey a given assumption, the imposition of that
assumption will contribute to the theoretical error in the extraction
of $\phi_s^{c\bar{c}s}$. That is,
\beq
\label{Edef}
E(1) \le E(2) \le ... \le E(8) ~, 
\eeq
where $E(N)$ is the theoretical error on $\phi_s^{c\bar{c}s}$ in the
scenario where $N$ assumptions are made.  (Note that, for each of
$N=8$ (the 1-amplitude method) and $N=1$ (the 2-amplitude method),
there is only a single scenario, but there are many possibilities for
$N=$2-7.)  It is true that the theoretical error associated with each
assumption is only $O(|A_{2}/A_{1}|)$. However, when one takes into
account the fact that multiple assumptions are involved, $E(8)$ is
quite a bit larger than $E(1)$. The theoretical error of the
2-amplitude method could well be an order of magnitude smaller than
that of the 1-amplitude method.

Note that this is not necessarily rigorously true. It is logically
possible that the entire theoretical error comes from the $D_{00}$
assumption, with all other assumptions being true. In this case all
the $E(N)$ of Eq.~(\ref{Edef}) are equal.  However, this is
exceedingly unlikely. It is far more reasonable to expect that each
assumption contributes to the theoretical error. In this case, fewer
assumptions lead to a smaller theoretical error, so that the
2-amplitude method is indeed an improvement over the 1-amplitude
method.

In fact, this can be tested experimentally by comparing the best-fit
values of $\phi_s^{c\bar{c}s}$ in the 1- and 2-amplitude methods --
their difference is $E(8) - E(1)$. (To be conservative, in computing
this difference, one should consider the full allowed ($1\sigma$)
range for $\phi_s^{c\bar{c}s}$.) If the theoretical error were due
solely to the $D_{00}$ assumption, one would have $E(8) - E(1) =
0$. If not, this proves that $E(8) \ne E(1)$.  This then quantifies
the improvement of the 2-amplitude method over the 1-amplitude method.

This type of analysis can be pushed further. Starting with the
1-amplitude method, we can relax an assumption by adding a new unknown
parameter. For example, suppose we allow $|{\bar A}_0| \ne |A_0|$, but
the other assumptions of Eq.~(\ref{assumptions}) are retained. Now,
when doing the fit to the data, there are 9 unknown parameters -- the
$|A_h|$ and $|{\bar A}_0|$ (5), the relative strong phases (3), and
$\phi_s^{c\bar{c}s}$. In addition, the theoretical expressions in
Table \ref{abcdtableA1} are modified; the correct expressions can be
found from Tables \ref{akcktable}-\ref{dktable}, but imposing the
assumptions that have been retained. The fit will give the preferred
values of $|{\bar A}_0|$ and $|A_0|$, so that we can see to what
extent, if any, the assumption that $|{\bar A}_0| = |A_0|$ is
violated. The fit will also give the preferred value of
$\phi_s^{c\bar{c}s}$. This can be compared with the value found in the
full 1-amplitude method, so this will give $E(8) - E(7)$, where the 7
assumptions of $E(7)$ are those that have been retained.

In fact, LHCb has performed an analogous analysis
\cite{LHCbBsmixing,privatecomm}. They allow for $|{\bar A}_h| \ne
|A_h|$, but take this difference to be helicity-independent. That is,
they assume that
\beq
\label{LHcbcond}
\frac{|{\bar A}_h|}{|A_h|} = \lambda ~~~~\forall~h ~,
\eeq
and add the unknown parameter $\lambda$ to the fit. This non-equality
has the effect of making nonzero, but helicity-independent, the $a_k$
and $c_k$ which vanish in Table \ref{abcdtableA1}. In the fit, the
idea is to see how much $\lambda$ deviates from 1. They find that this
deviation can be at most $\pm 5\%$, and leads to a deviation in
$\phi_s^{c\bar{c}s}$ of $\pm 0.02$ rad. We therefore have $E(8) -
E(7_{LHCb}) = 0.02$ rad, where here the 7 assumptions of $E(7_{LHCb})$
are those that have not been relaxed. (One way to write the LHCb
condition of Eq.~(\ref{LHcbcond}) is: $R_0 = R_\|$, $R_\| = R_\perp$,
$R_\perp = R_S$, $R_S = \lambda$, where $R_h \equiv |{\bar
  A}_h|/|A_h|$. Written in this way, we see that the assumption $R_S =
1$ of the 1-amplitude method has been relaxed, but 3 other assumptions
remain. This is indeed a 7-assumption scenario.)

Now, above we noted that $E(1)$, the theoretical error in the
2-amplitude method, cannot be reliably calculated and is
unmeasurable. However, because only one assumption is involved, this
error is $O(|A_{2}/A_{1}|)$. Similarly, we expect that $E(N) - E(N-1)$
is $O(|A_{2}/A_{1}|)$ since the two scenarios differ by one
assumption. One can therefore use the measured value of $E(N) -
E(N-1)$ to estimate $E(1)$. To give a concrete example, if one assumes
that the measured value of $E(8) - E(7_{LHCb}) = 0.02$ rad is a
``typical'' error, then one can infer that this is approximately
$E(1)$. Of course, this is a dangerous conclusion to draw.  The
expression for a theoretical error [e.g.\ Eq.~(\ref{D00eqn})]
generally also depends on a function of a strong phase. If this
function happens to be small for $E(7_{LHCb})$, $E(8) - E(7_{LHCb})$
will be atypically small. It is therefore more prudent to repeat the
above analysis by relaxing a different assumption in
Eq.~(\ref{assumptions}), and perhaps to do so more times. When it is
found that the value of $E(8) - E(7)$ associated with several
different assumptions is about the same size, that value can also be
assigned to $E(1)$. Of course, this prescription is clearly not
rigorous -- it is meant only to give a rough estimate at best.

We have argued that it is best to perform a fit with as many unknown
parameters as possible.  Ideally, 7 out of the 8 assumptions would be
relaxed, making a fit with 15 unknown parameters. This is the full
2-amplitude method. However, if a fit with 15 unknowns is not
possible, one can reduce the number of unknowns by adding assumptions
and incurring a larger theoretical error in $\phi_s^{c\bar{c}s}$. The
point is that one has a choice in the type of analysis that is
performed. In order to maximully reduce the effect of PP, one should
include as many free parameters in the fit as possible, i.e.\ make the
fewest possible assumptions.

Now, above we asked the hypothetical question: suppose a future
measurement finds $\phi_s^{c\bar{c}s} = 0.1 \pm 0.01~{\rm rad}$ (error
is statistical only).  This disagrees with the SM prediction
($2\beta_s = 0.03636 \pm 0.00170~{\rm rad}$) and therefore suggests
NP. Can we conclude that NP is indeed present when the theoretical
error on $\phi_s^{c\bar{c}s}$ is taken into account? We noted that it
is estimated theoretically that $|A_2 / A_1| = O(10^{-3})$. Suppose it
is a little larger, say 0.005.  And suppose that, in the 1-amplitude
method, the theoretical error is even larger, say $\simeq 0.04$,
because many assumptions are made. Thus, taking all errors into
account, one cannot conclude that NP is present. On the other hand, in
the 2-amplitude method, the theoretical error is still sufficiently
small ($\simeq 0.005$) that the presence of NP can be confirmed. We
therefore see that there are real advantages in the search for NP to
employing the 2-amplitude method.

\section{Applications}

If it is established that the measured value of $\phi_s^{c\bar{c}s}$
differs sufficiently from the SM prediction that one can conclude that
NP is present, this NP must be in the mixing.  One can also
independently test whether there is NP in the decay. One uses the
information about the magnitudes and relative strong phases of the
$A_h$ and ${\bar A}_h$ (14 experimental results $+$ 1 theoretical
input).  In order to do so, an assumption must be made regarding the
form of the decay amplitude. For example, suppose that it is assumed
that NP in the decay is not present, and that the amplitude has the
form of Eq.~(\ref{ABdefs}). Now there are 15 unknowns -- the
magnitudes and relative phases of $A_{1,h}$ and $A_{2,h}$ (the value
of the weak phase $\gamma$ can be taken from independent
measurements). The values of these unknowns can be determined from the
results on $A_h$ and ${\bar A}_h$. One can then test the theoretical
prediction that $|A_2 / A_1| = O(10^{-3})$. If there is a strong
discrepancy with this prediction, this might suggest the presence of
NP in the decay. (This is similar to the analysis of
Ref.~\cite{Fleischer99}.)

Suppose one assumes that the SM PP term is small, but that there
is NP in the decay. In this case, the decay amplitude takes the form
\beq
\label{NPdecay}
{\cal A}_s^{J/\psi \phi} = A_1 + e^{i\phi_{NP}} A_{NP} ~,
\eeq
where $\phi_{NP}$ is the (helicity-independent) NP weak phase. But now
there are 16 unknowns -- the magnitudes and relative phases of
$A_{1,h}$ and $A_{NP,h}$, and $\phi_{NP}$. These cannot all be
determined from the 15 magnitudes and relative strong phases of $A_h$
and ${\bar A}_h$. 

However, it has been argued that, to a good approximation, the NP
strong phases are all negligible \cite{DLphases}. Briefly, the logic
is as follows. Consider diagram $D$. At leading order its strong phase
is zero. But diagram $D'$ can rescatter into $D'_{resc}$, which has
the same Lorentz structure as $D$, and has a large strong phase. Note
that $|D'_{resc}/D'| \sim$ 5-10\%. We now have $D_{tot} = D +
D'_{resc}$, so that the diagram develops a strong phase. In the SM,
the dominant strong phases of the diagrams $C'$, $P'_{ct}$ and
$P'_{EW}$ are all generated via rescattering from the color-allowed
diagram $T'$. Since $|T'|$ is considerably larger than $|C'|$,
$|P'_{ct}|$ and $|P'_{EW}|$, the generated strong phases are
sizeable. On the other hand, the strong phase of a NP operator
$A_{NP}$ can only be generated by rescattering from itself:
$A_{NP,tot} = A_{NP} + A_{NP,resc}$. And since $|A_{NP,resc}/A_{NP}|
=$ 5-10\%, the NP strong phase is quite small.

In this case, the $A_{NP,h}$ helicity amplitudes all have the same
phase (0), and the total number of unknown parameters is 13 -- the
magnitudes of $A_{1,h}$ and $A_{NP,h}$ (8), the relative strong phases
(4), and $\phi_{NP}$. These can be determined from the 15 results, so
that we can measure the values of the NP parameters (given the initial
assumption).

Above, we have concentrated on $\Bsdecay$, but this 2-amplitude method
can be applied to other $\bs\to V_1 V_2$ decays. For example, consider
$\bs\to D_s^{*+} D_s^{*-}$. Its amplitude is given by
\bea
{\cal A}_s^{D_s^{*+} D_s^{*-}} & = & \lambda^{(s)}_c T' + (\lambda^{(s)}_t P'_t
+ \lambda^{(s)}_c P'_c + \lambda^{(s)}_u P'_u)
+ \frac23 \lambda^{(s)}_t P^{\prime C}_{EW} \nl
& = & \lambda^{(s)}_c (T' + P'_{ct} - \frac23 P^{\prime C}_{EW}) + \lambda^{(s)}_u (P'_{ut} - \frac23 P^{\prime C}_{EW}) \nn\\
&\equiv& A_1 + e^{i\gamma} A_2 ~.
\eea
Above, $T'$ and $P^{\prime C}_{EW}$ are the colour-allowed tree and
colour-suppressed electroweak penguin diagrams, respectively. As
before, the $A_2$ term represents the PP, and a rough estimate is
$|A_2 / A_1| = O(10^{-2})$. If the PP is neglected, the indirect CPA
in $\bs\to D_s^{*+} D_s^{*-}$ cleanly measures $\phi_s^{c\bar{c}s}$. However,
there is a theoretical error due to its neglect.

This error can be reduced as described above. One does a
time-dependent angular analysis of the decay, but allows for the
possibility of more than one contribution to the amplitude,
i.e.\ ${\bar A}_h \ne A_h$. The one difference is that, in $\bs\to
D_s^{*+} D_s^{*-}$, there is no $S$ helicity. Thus, the index $h$
takes only three values: $0, \|, \perp$. Consequently, there are only
six $h_k(t)$'s in Eq.~(\ref{AD:Bsphiphi}). But apart from this, the
analysis is unchanged -- $\phi_s^{c\bar{c}s}$ can be measured even in the
presence of PP, with a minimal theoretical error.

Another (slightly different) decay to which this method can be applied
is $\bs\to K^{*0} {\bar K}^{*0}$.  This is a pure penguin decay, whose
amplitude is given by
\bea
{\cal A}_s^{K^{*0} {\bar K}^{*0}} &=& \lambda^{(s)}_t P'_t + \lambda^{(s)}_c P'_c + \lambda^{(s)}_u P'_u \nn\\
         &=& \lambda^{(s)}_c P'_{ct} + \lambda^{(s)}_u P'_{ut} ~,
\label{KKamp}
\eea
where $P'_{ut} \equiv P'_u - P'_t$, $P'_{ct} \equiv P'_c - P'_t$.  It
is expected that $|P'_{ut}|$ and $|P'_{ct}|$ are of similar
size. Also, in contrast to $\Bsdecay$, these penguin amplitudes are
not OZI-suppressed.  Now, $|\lambda^{(s)}_u| = O(\lambda^4)$ and
$|\lambda^{(s)}_c| = O(\lambda^2)$, where $\lambda=0.23$ is the sine
of the Cabibbo angle. This suggests that the $\lambda^{(s)}_u P'_{uc}$
term can perhaps be neglected. However, for consistency, one must
neglect {\it all} $O(\lambda^4)$ terms. One of these is ${\rm
  Im}(\lambda^{(s)}_t)$, so that $\lambda^{(s)}_t$ is real.  But since
$\beta_s$ is related to ${\rm Im}(\lambda^{(s)}_t)/{\rm
  Re}(\lambda^{(s)}_c)$, it too vanishes in this limit. The net effect
is that, if one neglects $\lambda^{(s)}_u P'_{uc}$ (the ``penguin
pollution''), along with the other $O(\lambda^4)$ terms, the indirect
CPA in $\bs\to K^{*0} {\bar K}^{*0}$ vanishes because $\beta_s$ must
also be neglected.

That is, if one wishes to analyze the measurement of $\bs$-$\bsbar$
mixing in $\bs\to K^{*0} {\bar K}^{*0}$, the PP term must be kept in
the amplitude. Of course, there is now a theoretical error in this
result, related to the nonzero PP term. In order to estimate this error,
it has been suggested to measure the corresponding PP term in $\bd\to
K^{*0} {\bar K}^{*0}$ and relate it to $\bs\to K^{*0} {\bar K}^{*0}$
using flavour SU(3) \cite{CPSetc1,CPSetc2}.  (This method also applies to
$B^0_{d,s}\to K^0 {\bar K}^0$.) As always, there is an uncertainty due
to the unknown value of SU(3) breaking.

Once again, one can reduce the problems through a time-dependent
angular analysis of $\bs\to K^{*0} {\bar K}^{*0}$. The method is
identical to that in $\Bsdecay$, but with only three helicities ($0,
\|, \perp$), or six $h_k(t)$'s. Since both contributions to the
amplitude [Eq.~(\ref{KKamp})] are retained, one can extract $2\beta_s$
with a small theoretical error.

As is clear from the above, the 2-amplitude method for extracting
$2\beta_s$ with a minimal theoretical error in the presence of penguin
pollution can be applied to all $\bs/\bsbar \to V_1 V_2$ decays.  (In
the case of $\bs\to \phi \phi$, in which each $\phi$ is detected via
$\phi\to K^+ K^-$, it may be necessary to add a fifth helicity. This
corresponds to the case where both $K^+ K^-$ pairs are observed in
$S$-wave.) It can even be used for $\bs/\bsbar$ decays that are
governed by a $\btod$ transition. An example is $\bs\to J/\psi {\bar
  K}^{*0}$/$\bsbar \to J/\psi K^{*0}$ \cite{JpsiK*}. Here, in order
that both $\bs$ and $\bsbar$ can decay to the same final state, so as
to have an indirect CPA, one must consider the flavour-nonspecific
decay of the ${\bar K}^{*0}/K^{*0}$ to $\ks\pi^0$.

\section{Conclusions}

To summarize: in the standard model (SM), the weak phase of
$\bs$-$\bsbar$ mixing, $2\beta_s$, is expected to be very small. For
this reason, its measurement is of great interest, as a disagreement
with the SM prediction ($2\beta_s \simeq 0.036 ~{\rm rad}$) will
reveal the presence of new physics (NP).  The most common decay used
for measuring $2\beta_s$ is $\Bsdecay$.  Since this decay involves two
final-state vector mesons, a time-dependent angular analysis must be
used to disentangle the CP $+$ and $-$ final states. The LHCb
Collaboration has determined $\phi_s^{c\bar{c}s}$ (the measured value
of $2\beta_s$) to be small, in agreement with the SM, but still with
significant statistical and systematic errors.

In fact, there is also a theoretical error associated with this
measurement. $\Bsdecay$ is dominated by the colour-suppressed tree
diagram, but there are other contributions due to gluonic and
electroweak penguin diagrams. As these have a different weak phase
from the tree, their inclusion in the amplitude leads to a theoretical
error in the extraction of $2\beta_s$ from the data. They are
therefore often referred to as ``penguin pollution'' (PP). In the SM,
it is estimated that the ratio of the sizes of the two contributions
is $O(10^{-3})$, so that the PP is negligible. Still, there is some
uncertainty as to its exact size. This is important for the following
reason. Suppose a future measurement finds $\phi_s^{c\bar{c}s}$ to be
small, but in disagreement with the SM, even including the statistical
and systematic errors. Unless the theoretical error is under good
control, one cannot conclude that NP is present.

As $J/\psi$ and $\phi$ each have spin 1, the $\Bsdecay$ amplitude is
in fact 3 decays: the two spins are either parallel (helicity 0) or
transverse (two possibilities: helicity parallel ($\|$) or
perpendicular ($\perp$)). In addition, since $\phi \to K^+ K^-$, one
must also allow for the possibility that the observed $K^+ K^-$ pair
has relative angular momentum $l=0$ ($S$-wave -- helicity $S$). The
full $\Bsdecay$ amplitude therefore involves 8 amplitudes -- $A_h$ and
${\bar A}_h$ ($h=0,\|,\perp,S$). When PP in the amplitude is entirely
neglected (the ``1-amplitude method''), one has ${\bar A}_h = A_h$.
But this really corresponds to 8 theoretical assumptions -- both the
magnitudes and the phases of $A_h$ and ${\bar A}_h$ are taken to be
equal. While the theoretical error associated with each assumption may
be small, the net theoretical error due to the combination of multiple
assumptions is in fact quite a bit larger.

In this paper we propose a modification of the above angular analysis
in which the PP is included (the ``2-amplitude method''). The
amplitude is written in terms of all 8 $A_h$ and ${\bar A}_h$. We show
there are enough time-dependent measurements to extract the magnitudes
of $A_h$ and ${\bar A}_h$ and 6 of the 7 relative phases. For
the final relative phase, theoretical input is required. However, this
is not difficult to find. For example, we can assume that the
$A_0$-${\bar A}_0$ relative phase is zero. With this assumption,
$\phi_s^{c\bar{c}s}$ can be extracted. The key point here is that the
2-amplitude method only involves one assumption, and so the
theoretical error is quite a bit smaller than that of the 1-amplitude
method, which involves 8 assumptions. The 2-amplitude method is
therefore more promising in the search for NP through the measurement
of $\bs$-$\bsbar$ mixing. Indeed, if $\phi_s^{c\bar{c}s}$ differs from
the SM prediction, this points to NP in the mixing, and there is also
enough information to test for NP in the decay.

This same method can be applied to other $\bs/\bsbar \to V_1 V_2$
decays (sometimes with slight modification): e.g.\ $\bs\to D_s^{*+}
D_s^{*-}$, $\bs\to K^{*0} {\bar K}^{*0}$, $\bs\to \phi \phi$, etc. It
can even be used for $\bs/\bsbar$ decays that are governed by a
$\btod$ transitions, e.g.\ $\bs\to J/\psi {\bar K}^{*0}$/$\bsbar \to
J/\psi K^{*0}$, in which the ${\bar K}^{*0}/K^{*0}$ both decay to
$\ks\pi^0$.

\bigskip
\noindent
{\bf Acknowledgments}: We are extremely grateful to Olivier Leroy for
his numerous communications related to the subject of penguin
pollution. We thank Max Imbeault for helpful discussions. This work
was financially supported by NSERC of Canada (BB, DL), and by the
National Science Foundation under Grant No.\ NSF PHY-1068052 (AD).


\begin{thebibliography}{99}

\bibitem{Lenzetal} For example, see A.~Lenz, U.~Nierste, J.~Charles,
  S.~Descotes-Genon, A.~Jantsch, C.~Kaufhold, H.~Lacker and S.~Monteil
  {\it et al.},
  Phys.\ Rev.\ D {\bf 83}, 036004 (2011)
  [arXiv:1008.1593 [hep-ph]].

\bibitem{LHCbBsmixing} LHCb Collaboration, CERN-LHCb-CONF-2012-002,
  2012.

\bibitem{pdg}
K.~Nakamura {\it et al.}  [Particle Data Group],
  J.\ Phys.\ G {\bf 37}, 075021 (2010).

\bibitem{GHLR1} M.~Gronau, O.~F.~Hernandez, D.~London and
  J.~L.~Rosner, Phys.\ Rev.\ D {\bf 50}, 4529 (1994).

\bibitem{GHLR2} M.~Gronau, O.~F.~Hernandez, D.~London and
  J.~L.~Rosner, Phys.\ Rev.\ D {\bf 52}, 6374 (1995).

\bibitem{smallcorr1}
H.~Boos, T.~Mannel and J.~Reuter,
  Phys.\ Rev.\ D {\bf 70}, 036006 (2004)
  [hep-ph/0403085].

\bibitem{smallcorr2}
H.~-n.~Li and S.~Mishima,
  JHEP {\bf 0703}, 009 (2007)
  [hep-ph/0610120].

\bibitem{smallcorr3}
M.~Gronau and J.~L.~Rosner,
  Phys.\ Lett.\ B {\bf 672}, 349 (2009)
  [arXiv:0812.4796 [hep-ph]].

\bibitem{JpsiK*}
S.~Faller, R.~Fleischer and T.~Mannel,
  Phys.\ Rev.\ D {\bf 79}, 014005 (2009)
  [arXiv:0810.4248 [hep-ph]].

\bibitem{Psiphibetas}
R.~Aaij {\it et al.}  [LHCb Collaboration],
  Phys.\ Rev.\ Lett.\  {\bf 108}, 101803 (2012)
  [arXiv:1112.3183 [hep-ex]].

\bibitem{Swave}
S.~Stone and L.~Zhang,
  Phys.\ Rev.\ D {\bf 79}, 074024 (2009)
  [arXiv:0812.2832 [hep-ph]].

\bibitem{DDLR1} A.~S.~Dighe, I.~Dunietz, H.~J.~Lipkin and J.~L.~Rosner,
  Phys.\ Lett.\  B {\bf 369}, 144 (1996)
  [arXiv:hep-ph/9511363].

\bibitem{DDLR2} A.~S.~Dighe, I.~Dunietz and R.~Fleischer,
  Eur.\ Phys.\ J.\ C {\bf 6}, 647 (1999)
  [hep-ph/9804253].

\bibitem{DDLR3} B.~Tseng and C.~W.~Chiang,
  arXiv:hep-ph/9905338.

\bibitem{DFN} I.~Dunietz, R.~Fleischer and U.~Nierste,
  Phys.\ Rev.\ D {\bf 63}, 114015 (2001)
  [hep-ph/0012219].

\bibitem{CW1} N.~Sinha and R.~Sinha,
  Phys.\ Rev.\ Lett.\  {\bf 80}, 3706 (1998)
  [arXiv:hep-ph/9712502].

\bibitem{CW2} C.~W.~Chiang and L.~Wolfenstein,
  Phys.\ Rev.\  D {\bf 61}, 074031 (2000)
  [arXiv:hep-ph/9911338].

\bibitem{Datta:2012ky} For example, see
  A.~Datta, M.~Duraisamy and D.~London,
  Phys.\ Rev.\ D {\bf 86}, 076011 (2012)
  [arXiv:1207.4495 [hep-ph]].

\bibitem{Fleischer99} R.~Fleischer,
  Phys.\ Rev.\ D {\bf 60}, 073008 (1999)
  [hep-ph/9903540].

\bibitem{privatecomm} G. Raven and S. Hansmann-Menzemer, private communications.

\bibitem{DLphases}
A.~Datta and D.~London,
  Phys.\ Lett.\ B {\bf 595}, 453 (2004)
  [hep-ph/0404130].

\bibitem{CPSetc1}
M.~Ciuchini, M.~Pierini and L.~Silvestrini,
  Phys.\ Rev.\ Lett.\  {\bf 100}, 031802 (2008)
  [hep-ph/0703137 [HEP-PH]].

\bibitem{CPSetc2}
B.~Bhattacharya, A.~Datta, M.~Imbeault and D.~London,
  arXiv:1203.3435 [hep-ph].

\end{thebibliography}
\end{document}